\documentclass[prb,aps,twocolumn,showpacs,superscriptaddress,raggedbottom]{revtex4}

\usepackage{graphicx,bm}
\usepackage{dcolumn}
\usepackage{amsmath}

\def\mb#1{\mbox{\boldmath$#1$}}

\def\fig#1{Fig.~\ref{#1}}

\begin{document}
\title {Quantum transport in the presence of a finite-range
time-modulated potential}
\author{C. S. Tang}
\affiliation{Physics Division, National Center for Theoretical
Sciences, P.O. Box 2-131, Hsinchu 30013, Taiwan}
\author{C. S. Chu}
\affiliation{Department of Electrophysics, National Chiao Tung
University, Hsinchu 30010, Taiwan}

\begin{abstract}
Quantum transport in a narrow constriction, and in the presence of a
finite-range time-modulated potential, is studied.  The potential is
taken the form $V(x,t) = V_{0}\,\theta(x)\theta(a-x)\cos(\omega t)$,
with $a$ the range of the potential and $x$ the transmission
direction.  As the chemical potential $\mu$ is increasing, the dc
conductance $G$ is found to exhibit dip, or peak, structures when
$\mu$ is at $n\hbar\omega$ above the threshold energy of a subband.
These structures in $G$ are found in both the small $a$ ($a \ll
\lambda_{F}$) and the large $a$ ($a \gg \lambda_{F}$) regime.  The
dips, which are associated with the formation of quasi-bound states,
are narrower for smaller $a$, and for smaller $V_{0}$.  The
locations of these dips are essentially fixed, with small shifts
only in the case of large $V_{0}$.  Our results can be reduced to
the limiting case of a delta-profile oscillating potential when both
$a$ and $V_{0}a$ are small. The assumed form of the time-modulated
potential is expected to be realized in a gate-induced potential
configuration.
\end{abstract}


\maketitle

\section{INTRODUCTION}

Inelastic scattering processes in quantum transport have drawn
continuous attentions in the recent past.  One of the common model
used is to invoke a time-modulated potential, with a certain spatial
profile, to the
system.\cite{but82,coo85,jia90,azb91,bag92,fen93,roj93,wag94,chi94,gor95}
This model was first utilized by B\"{u}ttiker and
Landauer\cite{but82} who, in considering a time-modulated barrier,
attempted to extract a traversal time of a particle tunneling across
the barrier.  Later work from other groups have considered inelastic
scatterings in different configurations such as the double
oscillating barriers\cite{jia90}, the time-modulated barrier with a
delta profile\cite{azb91,bag92}, and the oscillating quantum well in
between two static barriers\cite{roj93,wag94}.  The model has also
been extended to incorporate the inelastic effects due to phonons by
introducing a time-modulated potential involving the phonon
operators.\cite{gel89,cai89,cai90,moh93}  In all these studies, the
features arising from the inter-mixing between the elastic
scattering, through a static potential, and the inelastic
scattering, through a time-varying potential, is emphasised.  In
particular, these studies have demonstrated, among others, the
interesting feedback effect of the inelastic scattering on the
elastic channel.  Even though the above model is appropriate only
for inelastic processes that preserve the phase coherence of the
transmitting particles, the model has practical importance because
the coherent inelastic scattering (CIS) can be realized, at least,
in the case when the time-modulated potential is well-specified.

\begin{figure}[bp]
\includegraphics[width=.4 \textwidth,angle=0]{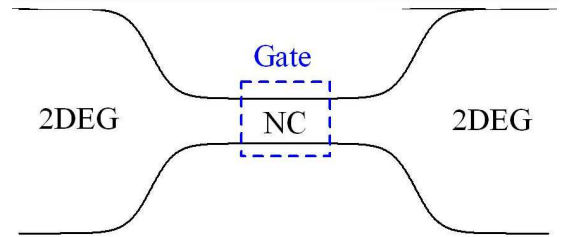}
\caption{Sketch of the gated QPC in which a NC is connected
adiabatically at each end to a 2DEG electrode. The gate induced a
finite-range time-modulated potential in the NC.} \label{fig1}
\end{figure}
A possible realization of CIS processes in nanostructures is
expected to be found in gate-controlled quantum point contacts
(QPC), as shown in \fig{fig1}.  Similar gate-induced potential
configuration has been suggested by Gorelik {\it et
al.},\cite{gor95} who considered microwave-induced effects on the
Josephson current through a narrow constriction (NC).  Their focus
is on the resonance of the microwave frequency with the energy
levels of the Andreev bound states formed in the NC, which has both
ends connected to superconducting electrodes.  For our purposes
here, a simple exhibition of the CIS is expected to be found readily
in a normal state gate-controlled QPC.   Recent development in the
split-gate technology has made possible the fabrication of such
gate-controlled QPC. \cite{wee88,wha88}  The split-gates, when
negative biased, define electrostatically a NC on a two-dimensional
electron gas (2DEG).  The dc quantum transport properties of these
QPC systems has been studied intensively. \cite{web92,buo93}  More
recently, there are growing interest in the time-dependent
properties, such as the effects of photons, in these QPC
systems.\cite{fen93,gri94,hu93,wys93,gor94,jan94} It is thus
legitimate to consider the quantum transport in a NC which is acted
upon by an additional, and ac biased, gate, as shown in \fig{fig1}.
This ac biased gate, which is different from the split-gates that
define the NC, induces on the NC a time-modulated potential.  The
scattering of the conduction electrons by this time-modulated
potential is both coherent and inelastic.

There is another reason why quantum transport in a NC in the
presence of a time-modulated, and gate-induced, potential is
interesting.  This is closely related to the density of states (DOS)
structures in the NC.  The energy levels in the NC are quantized
into one-dimensional subbands so that the DOS is singular at the
subband bottoms.  In the presence of attractive impurities, such
singularities in the DOS lead to dip structures in the dc
conductance $G$, as the chemical potential $\mu$ is increasing.
\cite{chu89,bag90,tek91,nix91,lev92,tak92,fai90,eug92,chu94}  The
dip structures occur when $\mu$ is just below a subband edge.
According to Bagwell,\cite{bag90} these dip structures are
associated with the formation of impurity-induced quasi-bound-states
(QBS).\cite{bag90}  The wavefunction at this energy $\mu$ and in
this subband is evanescent along the transmission direction. Hence
for the case of an attractive impurity, a QBS splits off from each
subband.\cite{bag90}  An electron originally in a propagating state
in other subband can thus be scattered elastically into and be
trapped by this QBS.  This gives rise to dip structures in $G$.

The QBS features in $G$ are found also when a point barrier
oscillates in a NC.\cite{bag92}  In this case, for a not-too-large
oscillation amplitude, the dc conductance $G$ exhibits dip, or peak
structures when $\mu$ is at $n\hbar\omega$ above a subband edge.
These structures correspond to the situation when the electrons can
make transitions, via inelastic processes, to the QBS just below the
subband edge.  That there is QBS, induces by the point oscillating
barrier, below each subband edge is demonstrated by Bagwell
\cite{bag92} from the energy poles in the current transmission
coefficients.  The imaginary part of the energy poles are negative,
which is consistent with the nature of the QBS.  The existence of
these QBS is again due to the singular DOS near a subband edge. It
is important to ask whether such QBS features in $G$ persist in the
case of a finite-range time-modulated (FRTM) potential.  This
question has not been addressed before and, if the QBS features did
persist in a FRTM potential, will have important implications to
time-dependent properties of QPC systems.  Furthermore, since the
potential is expected to be gate-induced, the problem is within
reach of the recent experimental capability.

In this paper, we simplify the problem by assuming the the
gate-induced potential in the NC to be represented by the form
$V_{0}\theta(x)\theta(a-x)\cos(\omega t)$, where $a$ is the range of
the potential and $x$ is the transmission direction.  Our
simplification is in replacing the smooth longitudinal potential
profile, of which the potential builts up within a longitudinal
distance of order $\lambda_{F}$, by an abrupt profile.  The
abruptness of the profile is expected to do nothing except
introducing additional multiple scatterings between the two abrupt
edges of the potential.  This results in  additional harmonics in
$G$.  Thus for the cases when the magnitude of these harmonics is
small, our results are expected to resemble qualitatively the
features in a smooth profile potential.  An explicit smooth-profile
consideration, however, is left to the further study.

Using this {\it small-harmonic-magnitude} criteria, we find
the QBS features in $G$ in both the small $a$ ($a \ll \lambda_{F}$)
and the large $a$ ($a \gg \lambda_{F}$) regime.  Our results can be
reduced to the limiting case of a delta-profile oscillating potential
when both $a$ and $V_{0}a$ are small.  In our calculation, the
inelastic scattering is solved nonperturbatively.  From
our results, we
note that even within the {\it small-harmonic-magnitude} criteria, the
inelastic scattering has to be treated beyond one sideband
approximation.  The sideband index $n$ labels those electrons which
net energy change is $n\hbar\omega$.

In Sec.\ II we present the formulation for the inelastic scattering
and the connection of the current transmission coefficients with the
conductance $G$.  In Sec.\ III we present
numerical examples illustrating the QBS features in a FRTM potential.
Finally, Sec.\ IV presents a
conclusion.

\section{THEORY}

In this section, the inelastic scattering problem is formulated and
the equations for the current transmission and
reflection coefficients are obtained.  The conductance $G$ is
then expressed in terms of these coefficients.

The QPC is modelled by a NC connecting adiabatically at each end to
a 2DEG.  Hence the transmission of the electrons into, or out of,
the NC region is adiabatic.\cite{gla88} The gate-induced potential
is assumed to affect only the NC region of the QPC.  Therefore we
need only to formulate the inelastic scattering in the NC region.
The NC is taken to have a quadratic transverse confinement potential
$\omega_{y}^{2} \, y^{2}$.  The gate-induced potential is taken the
FRTM form
\begin{equation}
V(x,t) = V_{0} \, \theta(x) \, \theta(a - x) \, {\rm \cos \/}
(\omega t),
\end{equation}
which connection with a smooth-profile potential has been discussed in
the previous section.

Choosing the energy unit $E^{*}=\hbar^{2} k_{F}^{2} / 2m^{*}$, the
length unit $a^{*}=1 / \! k_{F}$, the time unit $t^{*}=\hbar /
E^{*}$, and $V_{0}$ in units of $E^{*}$,
the dimensionless Schr\"{o}dinger
equation becomes
\begin{equation}
 \left[-\nabla^{2}+\omega_{y}^{2} \, y^{2}+V(x,t)
\, \right] \Psi(\mb{x},t) = i\frac{\partial}{\partial t} \,
\Psi(\mb{x},t).
\end{equation}
Here $k_{F}$ is a typical Fermi wavevector of the reservoir and
$m^{*}$ is the effective mass. The transverse energy levels are
quantized, with $\varepsilon_{n}= (2n+1) \, \omega_{y}$, and
$\phi_{n}(y)$ the wavefunction.  The FRTM potential is uniform in
the transverse direction and does not induce inter-subband
transitions, leaving the subband index $n$ unchanged.  Thus for a
$n$th subband electron incident along $\hat{x}$, and with energy
$\mu$, the scattering wavefunction can be written in the form
$\Psi_{n}^{+}(\mb{x},t) = \phi_{n}(y) \, \psi(x,t)$,
where\cite{fnt1}
\begin{widetext}
\begin{equation}
\psi(x,t) = \left\{ \begin{array}{lll}
e^{ik_{n}(\mu )x} \, e^{-i\mu t} +
\displaystyle{\sum_{m}} \,
r_{n}(m) \, \displaystyle{e^{-ik_{n}(\mu
+m\omega )x}} \, \displaystyle{e^{-i(\mu +m\omega)t}}, &\quad\mbox{if
$x<0$,} \\
\displaystyle{\sum_{p}} \,
\left[J_{p}(V_{0}/\!\omega ) e^{-ip\omega t}\right] \,
\displaystyle{\int
d\epsilon} \, \left[\tilde{A}_{n}(\epsilon )e^{ik_n(\epsilon )x} +
\tilde{B}_{n}(\epsilon )e^{-ik_n(\epsilon )x}\right] \,
e^{-i\epsilon t}, &\quad\mbox{if $0<x<a$,} \\
\displaystyle{\sum_{m}} \, t_{n}(m) \, e^{ik_{n}(\mu
+m\omega )x} \, \displaystyle{e^{-i(\mu +m\omega)t}} , &\quad\mbox{if
$x>a$,}     \end{array}
    \right.
\end{equation}
\end{widetext}
and $n$, $m$ are the final subband and sideband indices, respectively.
The effective wavevector for an electron with energy $\varepsilon$ and
in the $n$th subband is given by $k_{n}(\varepsilon) =
\sqrt{\varepsilon-(2n+1)\omega_{y}}$.  The sideband index $m$
corresponds to the net energy change of $m\hbar\omega$ for the
outgoing electrons.

It is very important to note that had the length of the NC been
infinite, and the range of the potential $V(x,t)$ were extended to
cover the entire NC, the longitudinal wavevector $k_{n}$ would be a
good quantum number so that no {\it real} transition could have
occurred.  However, as long as $V(x,t)$ has a finite range, $k_{n}$
is no longer conserved, and {\it real} transitions from
$k_{n}(\varepsilon)$ to $k_{n}(\varepsilon\pm m\omega)$ are
permitted for electrons traversing the potential.  Thus the
finiteness in the range of the time-modulated potential alone makes
possible the absorption of energy by the electrons for arbitrary
$\omega$.  This picture holds regardless of the range, long or
short, and the profile, abrupt or smooth, of the potential. The
mathematical statement of the above physical picture turns out
naturally, and given by Eq.(4), in the following.

The expressions for the reflection and the transmission coefficients
can be obtained from matching the wavefunctions, and their
derivatives, at the two ends of the FRTM potential.  For the
above matching to hold in all time, the integration variable
$\epsilon$ in Eq.(3) has to take on discrete values $\mu\pm m\omega$.
Hence we can write $\tilde{A}_{n}(\epsilon)$ and
$\tilde{B}_{n}(\epsilon)$ in the form
\begin{equation}
\tilde{F}_{n}(\epsilon ) = \sum_{m} \, F_{n}(m) \, \delta (\epsilon
-\mu -m\omega ),
\end{equation}
of which $\tilde{F}_{n}(\epsilon)$ refers to either
$\tilde{A}_{n}(\epsilon)$ or $\tilde{B}_{n}(\epsilon)$.  After
performing the matching, and the current reflection
coefficients $r_{n}(m)$ eliminated,
we obtain the equations relating $A_{n}(m)$, $B_{n}(m)$, and the
current transmission coefficients $t_{n}(m)$,
\begin{widetext}
\begin{equation}
t_n(m) = \sum_{m'} \, \left[A_{n}(m') \, e^{-iK^{-}_{n}(m,m')a}
+B_{n}(m') \, e^{-iK^{+}_{n}(m,m')a}\right] \, J_{m-m'}(V_{0}/\omega )
, \end{equation}
\begin{eqnarray}
k_{n}(\mu +m\omega ) \, t_n(m) & = &
\sum_{m'} \, k_{n}(\mu +m'\omega ) \, \left[A_{n}(m')
\, e^{-iK^{-}_{n}(m,m')a}
-B_{n}(m') \, e^{-iK^{+}_{n}(m,m')a}\right] \nonumber \\
 & &\hspace{7mm} \times J_{m-m'}(V_{0}/\omega ), \end{eqnarray}
and
\begin{equation}
2k_{n}(\mu ) \, \delta _{m0} =
\sum_{m'} \, \left[A_{n}(m') \, K^{+}_{n}(m,m')
+B_{n}(m') \, K^{-}_{n}(m,m')\right] \, J_{m-m'}(V_{0}/\omega ),
\end{equation}
\end{widetext}
where $K^{\pm }_{n}(m,m')=k_{n}(\mu +m\omega )\pm k_{n}(\mu
+m'\omega ).$  Equations (5)-(7) can be shown explicitly to reduce
to the corresponding equations for the delta-profile time-modulated
potential in the $a\rightarrow 0$ limit.\cite{chu95}

The zero temperature conductance is given by $G=(2e^2/h) \,
\sum_{n=0}^{N} \, G_{n}$, where $G_{n}=\sum_{m}G_{n}^{m}$, and $N+1$
is the number of propagating subbands in NC for the chemical
potential $\mu$.  The contribution to $G$ from electrons incident from
subband $n$ and transmit into sideband $m$ is given by $G_{n}^{m}$,
which is related to the transmission coefficients, given by
\begin{equation}
G_{n}^{m} = \left[ k_{n}(\mu+m\omega)/k_{n}(\mu)\right] \,
\mid t_{n}(m)\mid^{2}.
\end{equation}
Solving Eqs.(5)-(7), we obtain $t_{n}(m)$, $A_{n}(m)$, and $B_{n}(m)$,
using which the current reflection coefficient $r_{n}(m)$ can be
calculated,
\begin{equation}
r_{n}(m) = \sum_{m'} \, \left[A_{n}(m')+B_{n}(m')\right]
\, J_{m-m'}(V_{0}/\omega) \, -\delta_{m0}.
\end{equation}
We solve the coefficients $r_{n}(m)$, and
$t_{n}(m)$ exactly, in
the numerical sense, by imposing a large enough cutoff to the sideband
index.  The correctness of our procedure is checked against the
conservation of current condition, given by
\begin{equation}
\sum_{m} \, \frac{k_{n}(\mu+m\omega)}{k_{n}(\mu)} \,
\left[ \, |t_{n}(m)|^{2}+|r_{n}(m)|^{2}\right] = 1.
\end{equation}

\section{NUMERICAL RESULTS}

We calculate, in the following, the conductance $G$ of a NC acted upon
by a FRTM potential.
The FRTM potential does not induce
inter-subband transitions and so each occupied subband
contributes independently to the total conductance.  Thus it suffices
for our purposes here to present the conductance of only one subband,
which we take it to be the lowest one.

In this section, the behavior of $G$ with respect to the chemical
potential $\mu$ is studied.  Since $G$ depends also on the potential
range $a$, and the oscillating amplitude $V_{0}$, we present the
behavior of $G$ in four situations. Firstly, this $G$ behavior is
shown for $a$ fixed while varying $V_{0}$.  Secondly, the $G$
behavior is presented for a fixed $V_{d} = V_{0}a$, while varying
$a$.  This is to make connection with the delta-profile results, in
which the time-modulated potential is of the form $V_{d} \delta(x)
\cos(\omega t)$.  Thirdly, the $G$ behavior for $V_{0}$ fixed, while
$a$ varying, is presented.  The fourth situation is to compare the
$G$ behavior for different $\omega$.  In addition, we present the
time averaged spatial distribution for the scattering state, of
which the incident energy is very close to the QBS structure, and
the QBS is about to occur. Finally, we compare the one-sideband
approximation results with the nonperturbative results.
\begin{figure}[tbp]
\includegraphics[width=.4 \textwidth,angle=0]{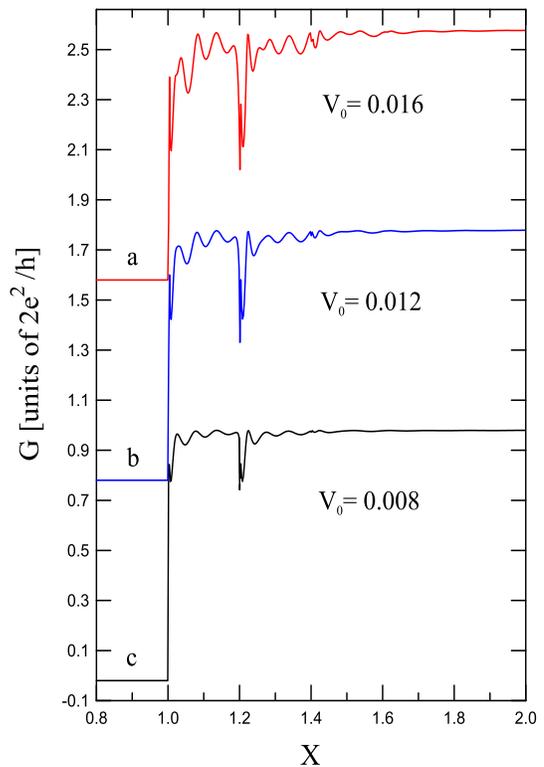}
\caption{Conductance $G$ as a function of $X$ for $a=150$, and
$\omega=0.014$. The potential oscillating amplitudes are: ``{\bf a}"
\ $V_{0}=0.016$, \ ``{\bf b}" \ $V_{0}=0.012$, \ ``{\bf c}" \
$V_{0}=0.008$. The curves are vertically offset for clarity.}
\label{fig2}
\end{figure}

In our numerical examples, the NC is taken to be that in a high
mobility GaAs-Al$_{x}$Ga$_{1-x}$As with a typical electron density $n
\sim 2.5 \times 10^{11}$ cm$^{-2}$, and $m^{*} = 0.067 m_{e}$.
Correspondingly, our choice of energy unit $E^{*} =
\hbar^{2}k_{F}^{2}/(2m^{*}) = 9$meV, length unit $a^{*} = 1/k_{F} =
79.6$ \AA, and frequency unit $\omega^{*} = E^{*}/\hbar = 13.6$THz.
We also take $\omega_{y} = 0.035$, such that the effective NC width is
of the order of $10^{3}$ \AA.  In the following, in presenting the
dependence of $G$ on $\mu$, it is more convenient to plot $G$ as a
function of $X$ instead, where
$X = [(\mu/\omega_{y})+1]/2$.  The integral value of $X$ is the number
of propagating channels.
\begin{figure}[tbp]
\includegraphics[width=.4 \textwidth,angle=0]{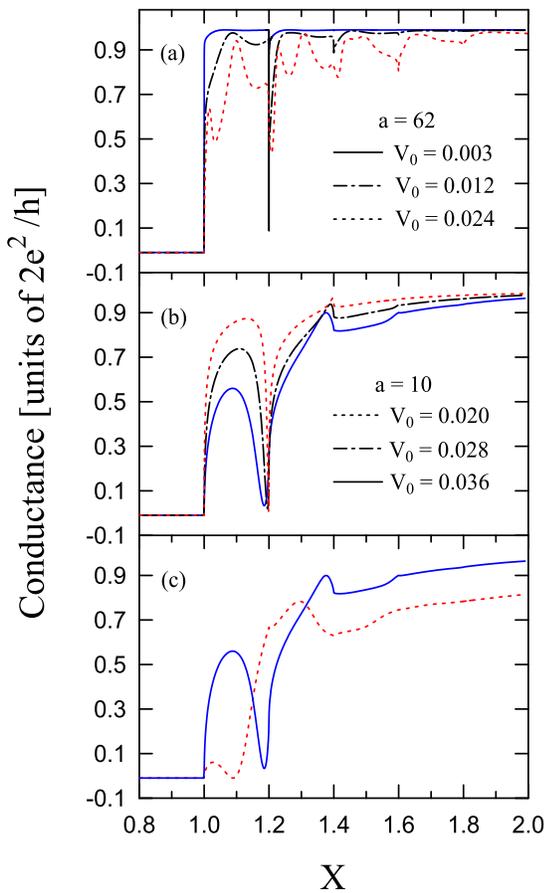}
\caption{$G$ as a function of $X$, for $\omega = 0.014$. (a) $a=62$,
$V_{0}=0.003, 0.012, 0.024$; (b) $a=10$, $V_{0}=0.020, 0.028,
0.036$; (c) FRTM potential result for $a=10, V_{0}=0.036$ (solid
curve). Delta-profile result for $V_{d}=0.36$ (dashed curve).}
\label{fig3}
\end{figure}

In Figs.\ 2 and 3, $G$ is plotted against $X$ for $a$ fixed, while
$V_{0}$ is varying.  The $a$ is fixed at 150, 62, and 10 in Figs.\
2, 3(a), and 3(b), respectively.  The frequency $\omega$ is taken to
be 0.014, which energy interval $\omega$ corresponds to an interval
$\Delta X = \omega/(2\omega_{y}) = 0.2$ on the ordinate.  The
threshold, or the subband edge, is at $X = 1$.  We note that in all
the cases shown, a major dip structure occurs at $X = 1.2$, which
corresponds to $X-\Delta X = 1$.  This is the QBS features because
the electron with energy at $X$ can make transition to the subband
edge by giving up an energy $\omega$.  We note also that, in
general, for larger $V_{0}$, the structures at $X = 1+N \, \Delta X$
are more evident.  These are the situations when the electron can
emit an energy of $N\omega$ and makes transition to the subband
edge.  The wavelength of the electron decreases as $X$ is
increasing.  The relation is given by $\lambda =
2\pi/\sqrt{2\omega_{y}(X-1)}$.  At the location of the first dip,
when $X=1.2$, we have $\lambda=53$.  Thus \fig{fig2} corresponds to
the case of a long potential range, with $a\simeq 2.8\lambda$, and
Fig.\ 3(b) corresponds to the case of a short potential range, with
$a\simeq 0.19\lambda$, near the occurrence of this first dip.
\begin{figure}[tbp]
\includegraphics[width=.4 \textwidth,angle=0]{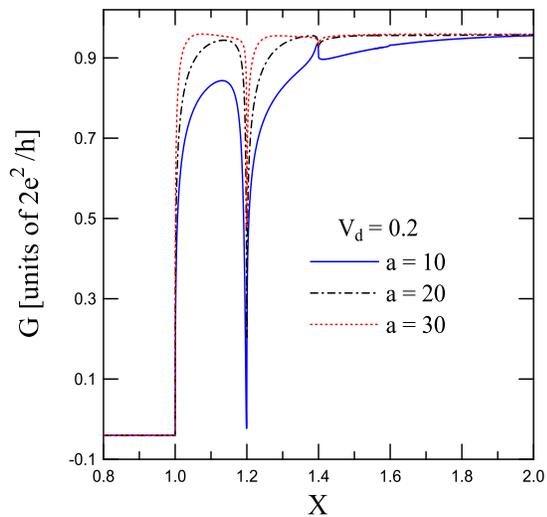}
\caption{$G$ as a function of $X$ for $V_{d}=0.2$, and $\omega =
0.014$. The parameter $a=10$ (solid curve); 20 (dash-dotted curve);
30 (dotted curve).} \label{fig4}
\end{figure}

Besides the QBS features, there are harmonic structures in
\fig{fig2} and in \fig{fig3}(a).  The structures are smaller for the
lower $V_{0}$. That these harmonics are associated with the multiple
scattering between the abrupt edges of the potential can be
identified from a resonance relation: $\lambda=2a/n$, with $n$ a
positive integer. Correspondingly, the harmonic peaks are at
$X_{n}=1+\Delta X_{n}$, with $\Delta
X_{n}=(n\pi/a)^{2}/(2\omega_{y})$.  According to the above estimate,
the first five harmonic peaks are at $X_{n}\simeq 1.006, 1.025,
1.056, 1.1,$ and $1.16$, which correspond quite reasonably to that
in Fig.\ 2.  However, for $X>1.2$, the harmonic peaks correspond
more closely to $X=1.2+\Delta X_{n}$.  This can be explained as
follows: the harmonics for $X>1.2$ are contributed mostly from those
electrons that give away an energy of $\omega$ so that the harmonics
at $X$ are very similar to that at $X-\omega$, or $X-0.2$ for our
cases.  In Fig.\ 2, we see that the harmonic amplitudes are
essentially smaller than the dip structure at $X=1.2$.  Using our
{\it small-harmonic-amplitude\/} criteria, the QBS features are
expected to be evident in a smooth-profile FRTM potential.  Similar
arguments can be applied to Fig.\ 3(a) to establish the harmonic
peak locations, but we do not repeat the detail here.  The harmonics
are essentially suppressed for $V_{0}=0.003$, with a very narrow dip
at $X=1.2$.  But at $V_{0}=0.012$, the harmonics are barely
emerging, and there is a new dip structure developing at $X=1.4$. At
$V_{0}=0.024$, the harmonic amplitudes become very large.  In the
short potential range case, as shown in Fig.\ 3(b), the dip
structure is narrower for the lower $V_{0}$ values. There is no
harmonic structures and the dip location shift only slightly towards
the lower $X$ region in the larger $V_{0}$ case. Even though at
$X=1.2$, we have $a\simeq 0.19\lambda$, the result does not
necessary reduce to the delta-profile result. We point out that only
the $V_{0}=0.02$ case in Fig.\ 3(b) corresponds to a delta-profile
result for $V_{d}=V_{0}a=0.2$.  To demonstrate this , we show in
Fig.\ 3(c) the $a=10$, and $V_{0}=0.036$ result (solid curve), and
compare with the delta-profile result for $V_{d}=0.36$ (the dotted
curve).  The two curves are very different. The most important
different is that the dip location in the delta-profile case is
shifted to $X\approx 1.1$ while the dip remains essentially
unchanged for the FRTM potential.
\begin{figure}[tbp]
\includegraphics[width=.4 \textwidth,angle=0]{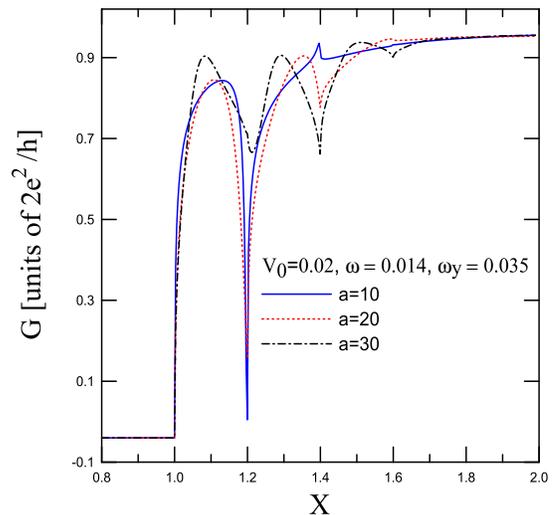}
\caption{$G$ as a function of $X$ for $V_{0}=0.02$, and $\omega =
0.014$. The parameter $a=10$ (solid curve); 20 (dash-dotted curve);
30 (dotted curve).} \label{fig5}
\end{figure}

To further establish the conditions when the FRTM potential result
reduces to a delta-profile result, we show in \fig{fig4} the FRTM
potential result for a small $V_{0}a = 0.2$ value, while varying
$a$. We recall that at $X=1.2$, $\lambda \simeq 53$.  The $a=10$
result (solid curve) is identical with the delta-profile result.
However, for $a=20$, and $30$, the FRTM results are different from
that of a delta-profile. As $a$ is increasing, the dip at $X=1.2$
becomes narrower and steeper, while the peak structure at $X=1.4$
becomes a very shallow dip.  This shows clearly that $a\ll \lambda$
must also be satisfied.  Thus the correct conditions for a FRTM
potential result to reduce to that of a delta-profile result are
$a\ll \lambda$, and $V_{0}a\ll 1$.
\begin{figure}[tbp]
\includegraphics[width=.4 \textwidth,angle=0]{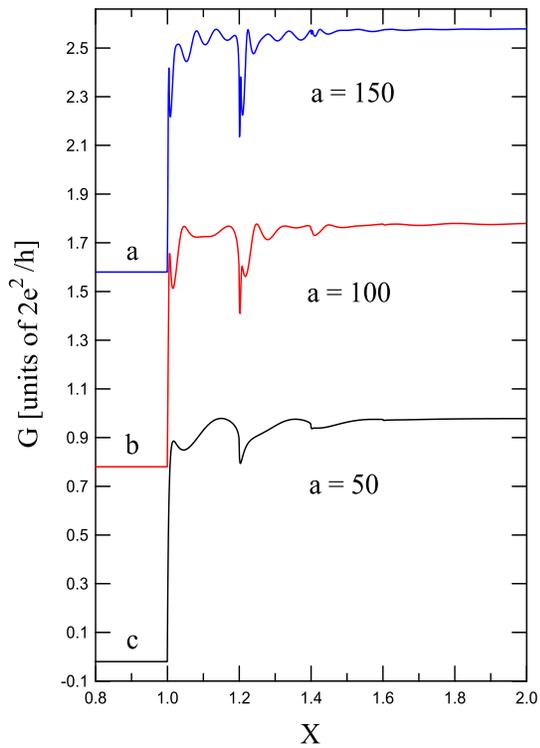}
\caption{$G$ as a function of $X$ for $V_{0} = 0.012$, and $\omega =
0.014$. The interaction ranges are: ``{\bf a}" \ $a=150$, \ ``{\bf
b}" \ $a=100$, \ ``{\bf c}" \ $a=50$. The curves are vertically
offset for clarity.} \label{fig6}
\end{figure}

In Figs.\ 5 and 6, we fix $V_{0}$ at 0.02, and varying $a$.  Fig.\ 5
is for the short potential range, with $a \leq \lambda$.  Within
this short $a$ regime, the dip at $X=1.2$ becomes wider and
shallower, as $a$ is increasing.  The dip at $X=1.4$, however,
becomes more pronounced as $a$ is increasing.  These features
indicate that high order processes such as the $2\omega$ processes
have become significant.  An electron that is trapped can then be
excited back out of the QBS. Thus we find the general trend that
when the $2\omega$ processes are significant, the first dip will
become shallower.  Fig.\ 6, on the other hand, is for the long
potential range cases.  Harmonics appear in the large $a$ regime.
We note that the $a=50$ result, which corresponds to the case
$a\simeq\lambda$, exhibits the emerging effects of the harmonics,
and the dip structures become very shallow.  For larger $a$, the dip
structures become larger again.
\begin{figure}[tbp]
\includegraphics[width=.4 \textwidth,angle=0]{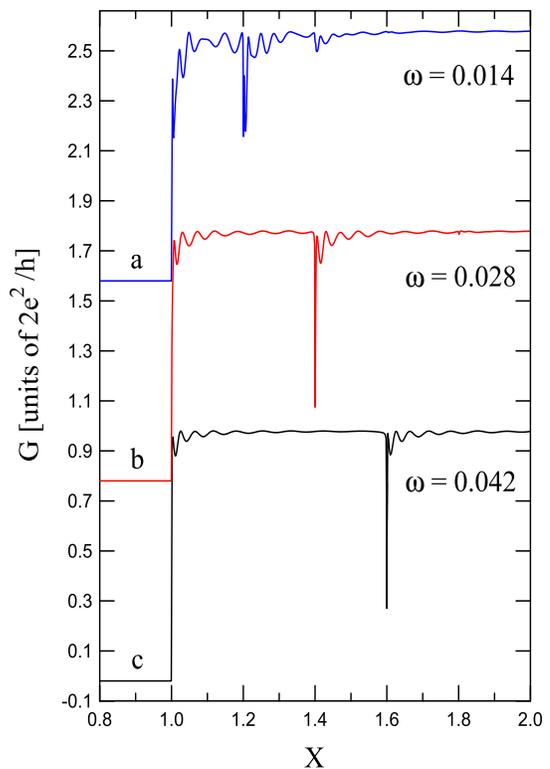}
\caption{$G$ as a function of $X$ for $a = 200$, and $V_{0} =
0.012$. The frequencies are: ``{\bf a}" \ $\omega =0.014$, \ ``{\bf
b}" \ $\omega =0.028$, \ ``{\bf c}" \ $\omega =0.042$. The curves
are vertically offset for clarity.} \label{fig7}
\end{figure}

\begin{figure}[tbp]
\includegraphics[width=.4 \textwidth,angle=0]{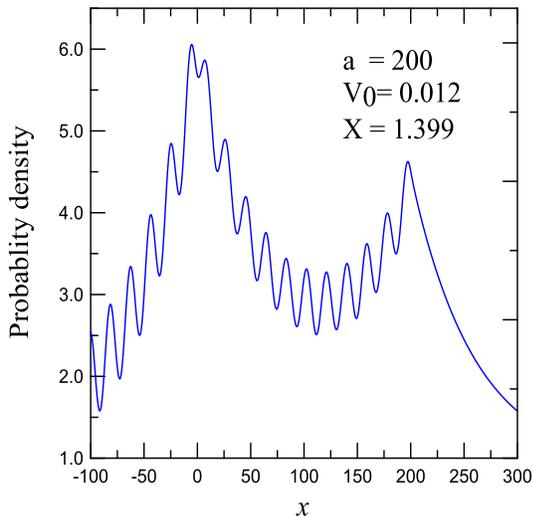}
\caption{Time-averaged probability density $\left< \mid
\psi(x,t)\mid^{2}\right>$ as a function of longitudinal position
$\it x$. The parameters are $a = 200, V_{0} = 0.012$, and $X =
1.399$.} \label{fig8}
\end{figure}
In \fig{fig7}, we present the case for $a=200$, $V_{0}=0.012$, and
varying $\omega$.  In curve {\bf a}, the dip structures are
subjected to the effect of the harmonics, since at the dip location,
the harmonic amplitude is not that small.  However, for curves {\bf
b}, and {\bf c}, the harmonic amplitudes are very small near the
location of the dip.  We point out also that the electron
wavelengths $\lambda$ near the dip structure in the curves {\bf b},
and {\bf c}, are 37.6, and 30.7, respectively. Thus, for example, in
curve {\bf c}, $a\simeq 6.5\lambda$, and we are in the very long
range regime.  The QBS features are still very clear.

\begin{figure}[tbp]
\includegraphics[width=.4 \textwidth,angle=0]{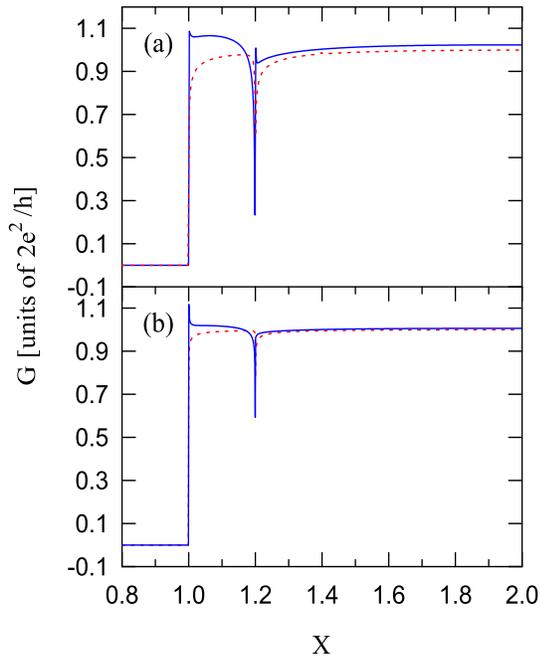}
\caption{$G$ as a function of $X$ for $a=10$, and $\omega =0.014$.
Nonperturbative result (dashed curve) and one-sideband approximation
result (solid curve). (a) $V_{0} = 0.01$, (b) $V_{0} = 0.005$.}
\label{fig9}
\end{figure}

In \fig{fig8}, we plot the time-average of the spatial probability
density for $a=200$, $V_{0}=0.012$, $\omega=0.028$, and $X=1.399$.
The dip location is at $X=1.4$.  Our choice of the parameters is
near that for the occurrence of the QBS.  The probability density
shows the evanescent nature of the trapped electron.  Furthermore,
the higher probability density near the two edges of the FRTM
potential shows that the QBS processes take place more frequently
within a distance of order $\lambda$ from the edges of the profile,
at which spatial variation in the potential occurs.

Finally, in \fig{fig9}, we present the one sideband approximation
results, together with our nonperturbative results. The deviation is
large in the $X<1.2$ region, which corresponds to the case when the
evanescent modes play an important role.  The deviation is larger
for larger $V_{0}$, as expected.  This shows that the one sideband
approximation is not appropriate and, in fact, the conservation of
current condition is usually not satisfied.

\section{CONCLUSION}

We have solved nonperturbatively the quantum transport in a NC, and in
the presence of an abrupt-profile FRTM potential.  The scattering
process is both inelastic and coherent.  We find QBS features in all
potential ranges, including the long as well as the short
range regime.  The dip structures associated with the QBS
occur when $\mu$ is at $m\hbar\omega$ above the threshold of a subband
edge.  We find also that the inelastic processes occur more likely
in the region when the potential profile varies spatially.
In addition, from our results which we have not
shown here,
we find that a one-sideband approximation, in general, violates the
conservation of current requirement.

We have presented arguments for the implications of our abrupt-profile
FRTM potential results to that of a smooth-profile FRTM potential.
We summarize and supplement our arguments, in the following, in
light of the similarity and the difference between the two potential
profiles.  The
abrupt-profile FRTM potential is similar to the smooth-profile FRTM
potential in that they both break the longitudinal translational
invariance.  This allows the electrons to absorb or emit energy in
units of $\hbar\omega$, for arbitrary $\omega$.  Consequently
the electrons can make transitions, via inelastic processes,
to the QBS just beneath a subband edge, giving rise in $G$ to dip
structures.  A conclusion from this similarity between the profiles is
that inelastic processes leading to QBS features are permitted in both
of the profiles.

These two potential profiles are different, however, in that
the abrupt-profile potential introduces additional
multiple scatterings between the two abrupt edges of the potential,
and gives rise to harmonics in $G$.
Due to these additional multiple scatterings, the electrons
effectively stay longer
within the abrupt-profile FRTM potential region than
within the smooth-profile FRTM potential region.  As a result, the QBS
features in the former potential profile might be perturbed, either
being enhanced or suppressed.  By proposing a {\it
small-harmonic-magnitude} criteria, we attempt to look at cases where
the perturbation from the harmonics to the QBS features is
expected to be small.
In these cases, the features of the two profiles are expected to be
similar qualitatively.

In conclusion, the CIS and the QBS features are found in a NC, acted
upon by an abrupt-profile FRTM potential.  The features are argued to
exhibit in the case of a smooth-profile FRTM
potential, in particular, and is expected to affect the time-dependent
properties of NC,
in general.  Further study, however, is needed to attain a better
understanding.

\acknowledgments{This work was partially supported by the National
Science Council of the Republic of China through Contract No.\
NSC85-2112-M-009-015.}


\begin{references}
\bibitem{fnt1} The wavefunction in the region $0<x<a$ is, by
construct, a solution of the time dependent Schr$\ddot{o}$dinger
equation, and has been proposed by Coon and Liu in Ref. 2, in their
study of the time dependent quantum well tunneling.
\end{references}
\end{document}